\documentclass[12pt,a4paper,hyper]{JHEP3}

\usepackage{amsmath}
\usepackage[pdftex]{graphicx}
\usepackage{multirow}
\usepackage{slashbox}

\newcommand{\eV}{{\rm eV}}

\newcommand{\MeV}{{\rm MeV}}
\newcommand{\GeV}{{\rm GeV}}
\newcommand{\TeV}{{\rm TeV}}

\newcommand{\Mpc}{{\rm Mpc}}

\newcommand{\pc}{{\rm pc}}

\newcommand{\muG}{\mu{\rm G}}

\newcommand{\nG}{{\rm nG}}

\newcommand{\ns}{{\rm ns}}

\newcommand{\Mpl}{M_{\rm Pl}}

\title{Lorentz Invariance Violation and Chemical Composition of Ultra High Energy Cosmic Rays}

\author{Andrey Saveliev$^{a}$, Luca Maccione$^{b}$,  Guenter Sigl$^{a}$ \\ $^{a}$II. Institut f\"ur Theoretische Physik, Universit\"at Hamburg,\\~Luruper Chaussee~149, 22761~Hamburg, Germany \\ $^{b}$DESY Theory Group, Notkestra{\ss}e~85, 22607 Hamburg, Germany  \\ E-Mail: \email{andrey.saveliev@desy.de, luca.maccione@desy.de, guenter.sigl@desy.de}}

\abstract{Motivated by experimental indications of a significant presence of heavy nuclei in the cosmic ray flux at ultra high energies ($\gtrsim 10^{19}~\eV$), we consider the effects of Planck scale suppressed Lorentz Invariance Violation (LIV) on the propagation of cosmic ray nuclei. In particular we focus on LIV effects on the photodisintegration of nuclei onto the background radiation fields. After a general discussion of the behavior of the relevant quantities, we apply our formalism to a simplified model where the LIV parameters of the various nuclei are assumed to kinematically result from a single LIV parameter for the constituent nucleons, $\eta$, and we derive constraints on $\eta$. Assuming a nucleus of a particular species to be actually present at $10^{20}$~eV the following constraints can be placed: $-3\times10^{-2} \lesssim \eta \lesssim 4$ for $^{56}$Fe, $-2\times10^{-3} \lesssim \eta \lesssim 3\times10^{-2}$ for $^{16}$O and $-7\times10^{-5} \lesssim \eta \lesssim 1\times10^{-4}$ for $^{4}$He, respectively.}

\keywords{cosmic ray theory, quantum gravity phenomenology, ultra high energy cosmic rays}

\preprint{DESY 10-224}

\begin{document}

\maketitle

\section{Introduction} 
\label{sec:Introduction}

Possible small deviations from the exact local Lorentz Invariance (LI) of general relativity have received a growing interest in recent years. On the theoretical side, hints of Lorentz Invariance Violation (LIV) were found in various approaches to Quantum Gravity (QG) \cite{Jacobson2006150, AmelinoCamelia:2008qg, Liberati:2009pf}. On the observational side, high energy astrophysics observations allowed to probe the consequences of LIV in various contexts \cite{Mattingly:2005re,Liberati:2009pf}. 

Lorentz symmetry breaking is not a necessary feature of QG, but it is clear that any possible LIV effect connected with the Planck scale could provide an observational window into QG. However, to directly observe phenomena connected with the Planck scale $\Mpl = 1.22\times10^{28}~\eV$ would require the center of mass energy of, e.g., a scattering process to be comparable to $\Mpl$. This is 15 orders of magnitude larger than what the LHC can probe with its design center of mass energy of $14~\TeV$. On the other hand, if we are testing LI specifically, then also non-LI quantities can be important. The energy of the particle in some frame, or a cosmological propagation distance are widely discussed examples. These quantities can be so large as to effectively offset the $\Mpl$ suppression to a physical observable, so that very small corrections are magnified. For this reason, they are called ``windows on QG''.

In order to correctly identify such ``windows on QG'' it is important to place them into a dynamical framework. A standard method is to study, within the context of EFT, a Lagrangian containing the standard model fields and all LIV operators of interest that can be constructed by coupling the standard model fields to new LIV tensor fields that have non-zero vacuum expectation values \cite{Colladay:1998fq,Myers:2003fd,GrootNibbelink:2004za,Bolokhov:2007yc,Mattingly:2008pw,Kostelecky:2009zp}. 

A word of caution should be made at this point. Lorentz invariance of physical laws relies on only few assumptions: the principle of relativity, stating the equivalence of physical laws for non-accelerated observers, isotropy (no preferred direction) and homogeneity (no preferred location) of space-time, and a notion of precausality, requiring that the time ordering of co-local events in one reference frame be preserved \cite{brown2005physical,Liberati:2001sd,Sonego:2008iu}. The above described procedure leads to breaking of the principle of relativity, but one may wonder if other possibilities exist that violate Lorentz invariance but preserve the principle of relativity. One such possibility is represented by the very special relativity framework \cite{Cohen:2006ky}, which corresponds to the break down of isotropy and is described by a Finslerian-type geometry~\cite{Bogoslovsky:2005cs,Bogoslovsky:2005gs,Gibbons:2007iu}. In this example, however, the new relativity group generators number less than the ten generators associated with Poincar\'e invariance. An example of a new relativity group different from Poincar\'e but with exactly ten generators is not known in commutative spacetimes. In non-commutative spacetimes, however, such a construction is possible and was termed ``doubly" or ``deformed" (to stress the fact that it still has 10 generators) special relativity, DSR~\cite{AmelinoCamelia:2011bm}. Unfortunately, the various DSR candidates  face in general major  problems regarding  their physical interpretation (they do not yield a low energy EFT, in general) and a working model is not yet available (see however \cite{AmelinoCamelia:2011bm} for recent attempts in new directions).
%

We now turn back to our preferred EFT approach. A generic result of the procedure described above is the presence of modified dispersion relations for particles, of the form
\begin{equation}
E^{2}-p^{2} = m^{2} + f(\vec{p},\Mpl; \mu)\;,
\label{eq:genericmdr}
\end{equation}
where $m$ is the particle mass, $E$ its energy, the function $f$ represents the QG contribution and can depend generically on the momentum $\vec{p}$, on $\Mpl$ and on some intermediate mass scale $\mu$. For simplicity we assume that only boost invariance is broken, while rotations are preserved (see \cite{Mattingly:2005re} for further comments on rotation breaking), so that $f$ depends on $p=|\vec{p}|$, rather than on $\vec{p}$. Moreover, at $p \ll \Mpl$ we can expand $f$ so that eq.~(\ref{eq:genericmdr}) reads
\begin{equation}
E^{2}-p^{2} = m^{2} + \sum_{n=0}^{N}\eta^{(n)} \frac{p^{n+2}}{\Mpl^{n}}\;,
\label{eq:mdrgeneral}
\end{equation}
where $\eta^{(n)}$ are constant dimensionless coefficients to be constrained. Terms with $n=0$ lead to strong deviations at small energies, while $n>0$ corresponds to high energy deviations from Lorentz symmetry. The latter situation is what we will discuss in this work.

Many of the parametrized LIV operators have been very tightly constrained via direct observations (see~\cite{Mattingly:2005re,Jacobson2006150, AmelinoCamelia:2008qg, Liberati:2009pf} for reviews). In particular, terms $n=2$, coming from dimension five and six CPT even LIV operators, have been recently directly\footnote{Notice that all these operators can be indirectly constrained by EFT arguments~\cite{Collins:2004bp}, as higher dimension LIV operators induce large renormalizable ones if we assume no other relevant physics enters between the TeV and $\Mpl$ energies.  Supersymmetry, however, is an example of new relevant physics that can change this argument.} constrained \cite{Maccione:2009ju} in the hadronic sector by exploiting ultra high energy cosmic ray observations performed by the Pierre Auger Observatory (PAO) \cite{PhysRevLett.101.061101}. Indeed, the successful operation of the PAO has brought UHECRs to the interest of a wide community of scientists and already allowed to test fundamental physics (in particular Lorentz invariance in the QED sector) with unprecedented precision \cite{PhysRevLett.100.021102,Maccione:2008iw,Galaverni:2008yj}.

The UHECR constraints \cite{Kluzniak:1999qq,Kifune:1999ex,Aloisio:2000cm,AmelinoCamelia:2000zs,Stecker:2004xm,GonzalezMestres:2009di,Scully:2008jp,Maccione:2009ju,Stecker:2009hj} rely on the behavior of particle reaction thresholds with LIV. What matters for threshold reactions in the presence of modified dispersion relations as in eq.~(\ref{eq:mdrgeneral}) is not the size of the LIV correction compared to the absolute energy of the particle, but rather the size of the LIV correction to the mass of the particles in the reaction.  Hence the LIV terms usually become important when their size becomes comparable to the mass of the heaviest particle. This criterion sets the presence of a critical energy $E_{cr}$ above which LIV effects are relevant in a given threshold reaction. If the LIV term scales with energy as $E^{n+2}$, then $E_{cr} \sim \left(m^{2}\Mpl^{n}\right)^\frac{1}{n+2}$ \cite{Jacobson:2002hd}. According to this reasoning, the larger the particle mass the higher is the energy at which threshold LIV effects come into play. 

UHECR constraints have relied so far on the hypothesis, not in contrast with any previous experimental evidence, that protons constituted the majority of UHECRs above $10^{19}~\eV$. Recent PAO observations \cite{Abraham:2010yv}, however, showed strong hints of an increase of the average mass composition with rising energies 
up to $E \approx 10^{19.6}~\eV$, although still with large uncertainties. Hence, experimental data suggests that heavy nuclei can possibly account for a substantial fraction of UHECR on Earth. Previous constraints need then to be revised. The purpose of the present work is to study the effects of LIV in the propagation of UHECR nuclei to establish more firm, albeit perhaps less stringent constraints.

This paper is composed as follows: in sec.~\ref{sec:LIV} we discuss our LIV framework, while in sec.~\ref{sec:UHE} we present general facts about propagation of UHECRs in the intergalactic medium. In sec.~\ref{sec:Photodisintegration} we deal with photodisintegration, the main process affecting the propagation of nuclear species, and develop a formalism to take possible LIV effects into account. The results of these calculations and some of their implications are finally presented in secs.~\ref{sec:Results} and \ref{sec:Conclusions}.

\section{Theoretical Framework} \label{sec:LIV} 

In order to study the phenomenological consequences of LIV induced by QG, the existence of a dynamical framework in which to compute reaction thresholds and rates is essential. We assume that the low energy effects of LIV can be parametrized in terms of a local EFT and, for simplicity, that rotation invariance be preserved. Therefore we introduce LIV by coupling standard model fields to a non-zero time-like vector.

We take the opportunity offered by the high energy involved in UHECR physics to focus our discussion on terms suppressed by two powers of $\Mpl$. 
The general discussion of this case can be found in \cite{Mattingly:2008pw,Kostelecky:2009zp}, where the structure of the LIV terms and the full set of modified dispersion relations can be found. Accordingly, the final dispersion relation for nuclei we want to further elaborate on is
\begin{equation}
E^2 =p^2+m^2 +\eta^{(2)}\frac{p^{4}} {\Mpl^{2}}~,
\label{eq:finaldisp}
\end{equation}
with $\eta$ independent of the helicity state of the nucleus, having assumed parity to be preserved. This is just a simplifying assumption that reduces the number of degrees of freedom we have to deal with in the following. Having already extracted the relevant powers of $\Mpl$,  $\eta$ should be of order $O(1)$ if the LIV effects are due to physics at the Planck scale. Therefore, any limit $|\eta| \ll 1$ is effectively a constraint on the assumption that QG physics occurs at the Planck scale. 

\subsection{Modified Dispersion Relations for heavy nuclei} \label{sec:DispRel}

The modified dispersion relations (MDRs) derived according to \cite{Mattingly:2008pw,Kostelecky:2009zp} are valid for elementary particles, such as electrons, neutrinos and quarks. Other MDRs hold for gluons, and the final MDR of a compound particle like a proton is a combination of all the LIV of the constituents. According to this approach \cite{Gagnon:2004xh}, a parton model needs to be assumed for protons and the net proton LIV is determined by the LIV terms for the partons along with the parton fraction at UHECR energies. However, we will not establish constraints on the bare parameters here, nor are we interested specifically in proton dispersion relations. Rather, we assume that each individual nucleus has its own independent MDR. 

Therefore, a nucleus with mass $A$ and charge $Z$ will have the following MDR
\begin{equation} \label{DispRel}
E^{2}_{A,Z} = p^{2}_{A,Z} + m^{2}_{A,Z} + \eta_{A,Z}^{(2)}\frac{p^{4}_{A,Z}}{M^{2}_{Pl}}\,,
\end{equation}
where $E_{A,Z}$ is the energy, $p_{A,Z}$ is the absolute value of the 3-momentum, $m_{A,Z}$ the mass and $\eta_{A,Z}^{(2)}$ the Lorentz-violating parameter of the nucleus. We assume that only nuclei have MDRs. Indeed, the target photons have too low energy for the LIV effects to be relevant in their dispersion relations.

Now we make a further simplification. Assuming a different $\eta_{A,Z}$ for each different nuclear species and isotope would lead to $O(100)$ free parameters to constrain, making it extremely hard to draw meaningful conclusions from the analysis of photodisintegration patterns. However, we can achieve a huge simplification by assuming that energy and momentum of the nucleus are the sum of energies and momenta of its constituents \cite{Jacobson:2002hd}. Using the fact that a nucleus is made of protons and neutrons and that protons and neutrons have almost the same mass, $m_{1}$, we have that the total energy and momentum of a nucleus $(A,Z)$ are given by  $E_{A} = \sum_{i} E_{i} \approx A E_{1}$ and $p_{A} = \sum_{i} p_{i} \approx A p_{1}$, respectively, where $E_{i}$ is the energy and $p_{i}$ is the momentum of the particular nucleon. With this approximation, the dispersion relation can be written as
\begin{equation}
E_{A}^{2} = \left( A E_{1} \right)^{2} = \left( A p_{1} \right)^{2} + \left( A m_{1} \right)^{2} + \frac{\eta}{A^{2}}\frac{\left( A p_{1} \right)^{4}}{\Mpl^{2}} = p^{2}_{A,Z} + m^{2}_{A,Z} + \frac{\eta}{A^{2}}\frac{p^{4}_{A,Z}}{\Mpl^{2}}\,.
\label{eq:mdr}
\end{equation} 
So now we have only one free parameter, $\eta$, for a nucleon, while for nuclei there are effective parameters of the form $\eta_{A} = \eta/A^{2}$. Although this is just a phenomenological model, we remark that it guarantees that the correct dispersion relations are recovered when dealing with macroscopic objects \cite{Jacobson:2002hd}, where the effects of QG physics should be suppressed. It might seem that the problem is now oversimplified. We remark, however, that the phenomenology of this model is already very rich for a first study of LIV effects on UHECR nuclei propagation, as it will be discussed later. We will instead leave to future work the study of a more general parameter space.



\section{Ultra High Energy Cosmic Rays} \label{sec:UHE}

The spectrum of Cosmic Rays (CRs) spans more than 12 decades in energy (ranging from $< 100~\MeV$ to $> 10^{20}~\eV$) with an impressively regular power-law shape
\begin{equation}
\frac{dN}{dE} \propto E^{-p(E)}\;.
\end{equation}
where the exponent $p(E)$ can be taken as a piecewise function, depending very little on the energy \cite{Gaisser:2006sf}.
 
One of the most puzzling problems of CR physics is related to their origin. Being charged particles, they are deviated while propagating in intergalactic and galactic magnetic fields, so that in general their arrival direction on Earth does not point back to their sources. As a consequence, the observed CR arrival directions are almost isotropically distributed. Only CRs with sufficiently high energy, $E \gtrsim 10^{20}~\eV$, are not deflected significantly in the $\nG$ intergalactic magnetic field, leading to a possible anisotropy, which was actually detected \cite{Cronin:2007zz,Abreu:2010zzj}. 

The Larmor radius of a relativistic particle with energy $E$, charge $Z$ and propagating in a magnetic field $B$ is 
\begin{equation}
r_{L} = \frac{E}{Z e B} \approx \frac{1.08}{Z} \;\pc  \left( \frac{E}{10^{15}~\eV} \right) \left( \frac{B}{\muG} \right)^{-1}\;.
\end{equation}
Accordingly, the $\muG$ galactic magnetic field is able to confine particles within the galactic disk up to $E\simeq Z\times10^{17}~\eV$, while CRs with higher energies cannot be confined and would produce a large anisotropy if they were accelerated in galactic source. Therefore, they are thought to have extragalactic origin.

The maximal energy of galactic CRs being dependent on the charge, we expect the actual end-point to the galactic CR spectrum to be determined by heavy nuclei, with the chemical composition becoming heavier and heavier with increasing energy. Experimental data in the energy range $10^{17}~\eV \lesssim E \lesssim 10^{18}~\eV$ are compatible with this expectation \cite{Kampert:2004rz}. At energy $E \gtrsim 10^{18}~\eV$, however, the chemical composition becomes again very light, being compatible with the presence of just protons \cite{hirescomp,augercomp}. 

Recently, the PAO measured the UHECR chemical composition up to $\sim 10^{19.6}~\eV$ \cite{Abraham:2010yv}. In apparent contrast with findings from other experiments \cite{hirescomp}, PAO results hint at the possibility that a significant fraction of UHECRs with energy $E\gtrsim 10^{19}~\eV$ might be constituted of heavy nuclei. Further observations which may confirm these results have been made by the Yakutsk EAS Array \cite{Glushkov:2007gd}. Because of limited statistics, it is not clear yet whether nuclei dominate the UHECR flux also above $10^{19.6}~\eV$. In the light of these results, it is then mandatory to study the propagation of heavy nuclei in the intergalactic medium (IGM). The IGM gas density being very small, hadronic interactions in the IGM can be neglected and we can focus on interactions of heavy nuclei with photon background fields, in particular the CMB and infrared fields. In this respect, the main interaction channels are the following \cite{ApJ.205.638}:
\begin{description}
 \item[Compton interactions] Inverse Compton scattering can be described as the process $^{A}_{Z}N +\, \gamma \rightarrow \, ^{A}_{Z}N +\, \gamma'$, i.e.~the UHE nucleus scatters on a (low energy) photon and transfers some of its energy to it. The energy loss is quite small but it is remarkable that this reaction, in comparison to the others, has no threshold.

\item[Electron-positron pair production] Pair production, $^{A}_{Z}N +\, \gamma \rightarrow \, ^{A}_{Z}N +\, e^{-} +\, e^{+}$, only dominates for nucleus energy below $10^{20}$~eV \cite{ApJ.205.638}. In the case of black body radiation at temperature $T$ the inverse attenuation length can be written as
\begin{equation}
\lambda^{-1}_{att} = \frac{1}{E} \frac{dE}{dx} = \frac{\alpha r^{2}_{e} Z^{2} \left( m_{e} k_{B} T \right)^{2}}{\pi p} f\left( \frac{m_{e}}{2 \gamma_{CR} k_{B} T} \right).
\end{equation} 
Here, $\alpha$ is the fine-structure constant, $T$ is the temperature of the corresponding photon field, $r_{e}$ is the classical radius of an electron and $m_{e}$ its mass. The function $f$ is given in \cite{PhysRevD.1.1596}. Since the reaction occurs frequently throughout the propagation and the energy loss in one interaction is of the order $m_{e}/m_{p} \approx 10^{-3}$, it may be treated as a continuous energy loss process. 

\item[Pion production] Pion production is a photohadronic process of the basic type $^{A}_{Z}N + \gamma \rightarrow ^{A'}_{Z'}N' + X + \pi^{0,\pm}$. Its cross section for single nucleons is well-known and can be found in \cite{PASA.16.160}. For nuclei usually a simple model for a combined cross section,
\begin{equation}
\sigma_{A,Z} = Z \sigma_{p} + (A - Z) \sigma_{n},
\end{equation}
$\sigma_{p}$ and $\sigma_{n}$ being the cross sections for a proton and a neutron, respectively, is used. In the LI case it begins to become important in comparison to other energy losses for energies which lie beyond the energy scales discussed here. In some sense, this process can be viewed as a kind of photodisintegration, in which the energy available is large enough as to trigger the formation of pions.

 \item[Photodisintegration] This is the most important interaction channel for changes of the chemical composition, also for the LIV case, hence we will describe it in more detail in sec.~\ref{sec:Photodisintegration}.
\end{description} 

Photodisintegration is the most relevant process for the propagation of heavy nuclei above $10^{19}~\eV$ \cite{ApJ.205.638}. In principle, also LIV effects on Compton scattering and on pair production should be considered. However, given that these processes are only relevant at low energy, we do not expect strong constraints coming from their analysis. Therefore, we will neglect them and we will focus on photodisintegration in the following.

The radiation fields relevant for our analysis are the Cosmic Microwave and Infrared Backgrounds (CMB and CIB). The ambient photon density per photon energy is given by
\begin{equation} \label{nCMB}
\frac{dN_{CMB}(\epsilon)}{dV d\epsilon} = \frac{1}{\pi^{2}} \frac{\epsilon^{2}}{e^{\frac{\epsilon}{k_{B} T}}-1},
\end{equation} 
$T=2.725\,$K, for the CMB, i.e.~it follows Planck's law of black-body radiation for the corresponding temperature. In the case of Cosmic Infrared Background the energy dependence is more complicated \cite{Ann.Rev.Astr.Astroph.39.1.249}.

\section{Photodisintegration of UHE Nuclei and its LIV Modifications} \label{sec:Photodisintegration}

Photodisintegration is the main process causing a change of a nucleus species during the propagation in the intergalactic medium. It may be described as a process consisting of two steps: Photo-absorption by the nucleus forming an excited compound state which then decays with the emission of (at least) one nucleon \cite{SalamonStecker_Photodisintegration}.

The process occurs with the so called Giant Dipole Resonance (GDR), for photon energies up to around 30~MeV in the rest frame of the nucleus, while at higher energies the quasi-deuteron process, baryon resonances and finally, at extremely high energies, (multi)fragmentation, i.e.~the decay of the initial nucleus into several large fragments, dominates \cite{Rachen_CosmicRaysProp}.

We are now going to consider a reaction of the form
\begin{equation}
^{A}_{Z}N + \gamma \rightarrow \, ^{A'}_{Z'}N' +\, ^{B}_{W}N''\;,
\end{equation}

where $A'=A-B$ and $Z'=Z-W$.

\subsection{Threshold Computations}

One of the most well known consequences of LIV in MDRs is to modify the threshold structure of scattering processes.

For the purpose of this threshold computation, we will consider, instead of an MDR like in eq.~(\ref{eq:mdr}), a generic MDR like

\begin{equation}
E_{A,Z}^{2} = p_{A,Z}^{2} + m_{A,Z}^{2} + \eta^{(n)}_{A,Z}\frac{p_{A,Z}^{n+2}}{\Mpl^{n}}\;,
\end{equation}

where the case of eq.~(\ref{eq:mdr}) is recovered by setting $n=2$ and $ \eta^{(2)}_{A,Z} = \eta/A^{2}$.

The energy-momentum conservation equation in the threshold configuration \cite{Jacobson:2002hd} reads
\begin{equation} \label{EnergyCons1}
(E_{A,Z}+\epsilon)^2-(\vec{p}_{A,Z}+\vec{p}_\gamma)^2 = (E_{A',Z'}+E_{B,W})^2-(\vec{p}_{A',Z'}+\vec{p}_{B,W})^2\,,
\end{equation}
where $E_{A,Z}$ and the other energy values are given by eq.~(\ref{DispRel}) while $\epsilon$ and $\vec{p}_\gamma$ are the energy and the momentum of the background photon. The threshold conditions are head-on collision of the incoming particles and parallel outgoing momenta \cite{Mattingly:2002ba}, i.e.~$\vec{p}_{A,Z} \vec{p}_\gamma = - p_{A,Z} p_\gamma = - p_{A,Z} \epsilon$ and $\vec{p}_{A',Z'} \vec{p}_{B,W} = p_{A',Z'} p_{B,W}$. Due to the large energy of the nuclei the products are emitted almost parallel to the direction of the initial nucleus, i.e.~momentum conservation requires $p = p_{A,Z} \approx p_{A',Z'} + p_{B,W} \approx (1-y) p + y p$ ($0 < y < 1$), with $y$ being the inelasticity of the reaction. After some algebra, the main threshold equation reads
\begin{equation} \label{ThresholdEquation}
\frac{\eta_{A,Z}^{(n)} - \eta_{A',Z'}^{(n)} (1-y)^{n+1} - \eta_{B,W}^{(n)} y^{n+1}}{M^{n}_{Pl}} p^{n+2} + 4 \epsilon p- \left(\frac{m^{2}_{A',Z'}}{1-y}+\frac{m^{2}_{B,W}}{y}-m^{2}_{A,Z}\right)
 = 0\,.
\end{equation}
To find a lower threshold, eq.~(\ref{ThresholdEquation}) has to be solved for $p$ and minimized with respect to $y$. However, in the case of LIV eq.~(\ref{ThresholdEquation}) may have two real and positive solutions, in which case there are two thresholds - a lower ($\underline{p}_{thr}$) and an upper ($\overline{p}_{thr}$) one (which has to be found by \textit{maximizing} $p$ with respect to $y$) \cite{Mattingly:2002ba}. If instead there are no real and positive solutions, then the reaction cannot take place at all.
\begin{figure}[tbp]
  \centering
 \fbox{\includegraphics[width=0.98\textwidth]{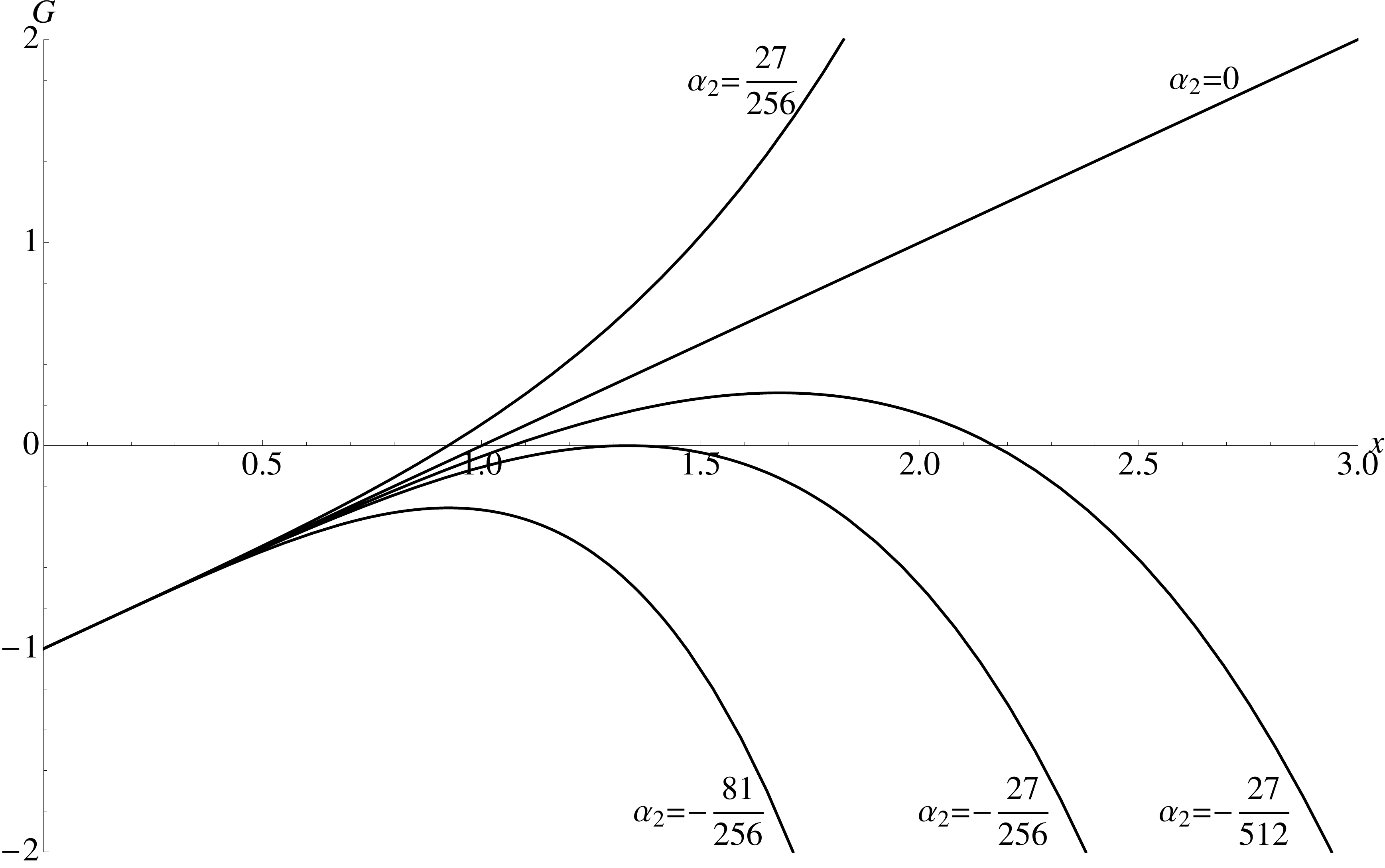}}
  \caption{$G(x,y)$ for $n=2$ and different values of $\alpha_{2}$. The points of intersection with the $x$ axis are the thresholds. The following configurations are given: For $\alpha_{2}=\frac{27}{256}$ and $\alpha_{2}=0$ (LI) there is one (lower) threshold; $\alpha_{2}=-\frac{27}{512}$ gives a lower and an upper threshold; $\alpha_{2}=-\frac{27}{256}$ is the tangential value $\alpha^{tang}_{2}$; for $\alpha_{2}=-\frac{81}{256} < \alpha^{tang}_{2}$ the reaction is kinematically forbidden.}
  \label{fig:Galpha}
\end{figure}
Dividing eq.~(\ref{ThresholdEquation}) by its last term and substituting 
\begin{equation} \label{alphaPoles1}
\begin{split}
\alpha_{n}(y) &= \frac{\eta^{(n)}_{A,Z} - \eta^{(n)}_{A',Z'} (1-y)^{n+1} - \eta^{(n)}_{B,W} y^{n+1}}{M^{n}_{Pl} (4\,\epsilon)^{n+2}} \left(\frac{m^{2}_{A',Z'}}{1-y} + \frac{m^{2}_{B,W}}{y} - m^{2}_{A,Z}\right)^{n+1}\\
x &= \frac{4 \epsilon}{\frac{m^{2}_{A',Z'}}{1-y} + \frac{m^{2}_{B,W}}{y} - m^{2}_{A,Z}} p\,,
\end{split}
\end{equation}
we obtain a polynomial of the order $n+2$:
\begin{equation} \label{ThresholdDisint}
G(x,y) := \alpha_{n}(y) x^{n+2} + x - 1 = 0\,,
\end{equation}
shown in fig.~\ref{fig:Galpha}. The number of solutions now depends on $\alpha_{n}(y)$. 

The region in which the process is generically allowed to take place in some energy window is limited by the tangential configuration ($G(x,y)=0$ and $\partial_{x}G(x,y)=0$ for $x>0$), given by
\begin{equation} \label{TangValues}
\alpha^{tang}_{n} = -\frac{(n+1)^{n+1}}{(n+2)^{n+2}},\qquad x^{tang}_{n} = \frac{n+2}{n+1}.
\end{equation} 
Using this and the structure of $\alpha_{n}(y)$, we obtain the results summarized in Tab.~\ref{tab:LimitsAlphaFinal_short}.

\begin{table}[tbp]
\centering
\begin{tabular}{|cc||c|c|c|c|c|}
\hline
\multirow{2}{*}{}                                      &      & $\eta^{(n)}_{A,Z}<\eta^{(n)}_{B,W}$ & \multicolumn{3}{c|}{$\eta^{(n)}_{A,Z}=\eta^{(n)}_{B,W}$} &
$\eta^{(n)}_{A,Z}>\eta^{(n)}_{B,W}$ \\
                                                       &      &                                     & $<0$ & $=0$ & $>0$ & \\
\hline
\hline
$\eta^{(n)}_{A,Z}<\eta^{(n)}_{A',Z'}$                  &      & PD constr.                          & PD constr. & PD constr. &       & \\
\hline
\multirow{3}{*}{$\eta^{(n)}_{A,Z}=\eta^{(n)}_{A',Z'}$} & $<0$ & PD constr.                          & PD constr. & excl.      & excl. & \\
\cline{2-7}
                                                       & $=0$ & PD constr.                          & excl.      & LI         & excl. & \\
\cline{2-7}
                                                       & $>0$ &                                     & excl.      & excl.      &       & \\
\hline
$\eta^{(n)}_{A,Z}>\eta^{(n)}_{A',Z'}$                  &      &                                     &            &            &       & \\
\hline
\end{tabular}
\caption{For some combination of parameters from the region labeled by ``PD constr.'' a kinematical configuration for photodisintegration to take place cannot be found so the reaction is forbidden and therefore it is possible to set constraints. Inside the white region the reaction is always possible at some energy, while the cases marked by ``excl.'' are excluded since the corresponding combination of conditions for $\eta_{A,Z}^{(n)}$ is not possible.}
\label{tab:LimitsAlphaFinal_short}
\end{table}

In general, constraints on the parameters can be placed by imposing that the reaction be allowed \textit{and} happen between some lower and upper threshold energy which are compatible with experimental results. 

\subsection{Propagation Lengths} \label{sec:PropLen}

There are two main quantities to characterize the effects of photodisintegration on the propagation of UHE nuclei. On the one hand we have the mean free path (the distance between two successive reactions)  \cite{PhysRev.180.1264}
\begin{equation} \label{MFP}
\lambda_{MFP}^{-1}\left(p\right) = \frac{m_{A,Z}^{2}}{2 \beta p^{2}} \int_{\epsilon_{thr}}^{\infty} \frac{d\epsilon}{\epsilon^{2}} \frac{dN(\epsilon)}{dV d\epsilon} \int_{\underline{\epsilon}'}^{\overline{\epsilon}'} d\epsilon' \left(\epsilon' -  \frac{1}{2 m_{A,Z}} \eta^{(n)}_{A,Z} \frac{p^{n+2}}{\Mpl^{n}} \right) \sigma(\epsilon')\,,
\end{equation}
with $\displaystyle \underline{\epsilon}' = \frac{p}{m_{A,Z}}(1-\beta) \epsilon + \frac{1}{2 m_{A,Z}} \eta^{(n)}_{A,Z} \frac{p^{n+2}}{\Mpl^{n}}$ and $\displaystyle \overline{\epsilon}' = \underline{\epsilon}' + 2\beta\frac{p}{m_{A,Z}}\epsilon$. 
On the other hand we consider the attenuation length, i.e.~the mean distance over which the particle energy is reduced by a factor $1/e$ with respect to its initial value, \footnote{It is not possible to define properly an attenuation length for a process that implies the disruption of the initial particle, such as photodisintegration is. Here we assume that we can group different nuclear species in such a way that the particle ``identity'' is not changed after a few $\lambda_{MFP}$. This is reasonable given the poor experimental accuracy in the determination of the mass number and charge of an UHECR nucleus. Moreover, assuming that only one or two nucleons are removed and that the nucleus Lorentz factor is almost constant, the average cross section changes smoothly with the mass and the continuous mass loss approximation can be valid \cite{Rachen_CosmicRaysProp}.} 
\begin{equation} \label{AttPath}
\lambda_{att}^{-1}\left(p\right) = -\frac{1}{E} \frac{dE}{dx} = \frac{m_{A,Z}^{2}}{2 \beta p^{2}} \int_{\epsilon_{thr}}^{\infty} \frac{d\epsilon}{\epsilon^{2}} \frac{dN(\epsilon)}{dV d\epsilon} \int_{\underline{\epsilon}'}^{\overline{\epsilon}'} d\epsilon'  \left(\epsilon' -  \frac{1}{2 m_{A,Z}} \eta^{(n)}_{A,Z} \frac{p^{n+2}}{\Mpl^{n}} \right) \bar{y}(\epsilon',p) \sigma(\epsilon')\,.
\end{equation}
In equations (\ref{MFP}) and (\ref{AttPath}) $\epsilon_{thr}$ is the threshold energy for the background photon obtained from eq.~(\ref{ThresholdEquation}), $dN(\epsilon)/dV d\epsilon$ is the ambient photon density, $\bar{y}(\epsilon',p)$ is the inelasticity as computed following \cite{PhysRevD.67.083003} and $\beta = \partial E / \partial p$. The cross section $\sigma(\epsilon')$ used here is taken from \cite{ApJ.205.638}, but modified by shifting the cutoff to $\epsilon_{max}'=10\,\GeV$.
The integration limits in equations (\ref{MFP}) and (\ref{AttPath}) are computed following the same procedure as \cite{Maccione:2009ju,PhysRevD.67.083003}.

For the single parameter model described by eq.~(\ref{eq:mdr}), $\bar{y}(\epsilon',p)$ is almost independent of $\eta$. Therefore it is enough to consider only  the mean free path $\lambda_{MFP}$ of eq.~(\ref{MFP}) from which the corresponding attenuation length can be obtained dividing by $y_{thr,LI} = m_{B,W}/(m_{A',Z'} + m_{B,W})$. In the general case in which we do not follow the approximation given in eq.~(\ref{eq:mdr}), the inelasticity, and with it the attenuation length, may be dramatically different from the mean free path, and therefore has to be considered in more detail.

\subsection{Spontaneous Decay}

Spontaneous decay of nuclei is a process described by $^{A}_{Z}N \rightarrow \, ^{A'}_{Z'}N' +\, ^{B}_{W}N''$. In this work we are mainly interested into decays in which there is \textit{one} heavy daughter nucleus and possibly other much lighter fragment(s). Some nuclei can decay spontaneously even without LIV. This occurs mainly for nuclei with either proton or neutron excess emitting (most probably) single nucleons (e.g.~$\text{$^{5}_{2}$He} \rightarrow \text{$^{4}_{2}$He} + \text{$^{1}_{0}$n}$ or $\text{$^{9}_{5}$B} \rightarrow \text{$^{8}_{4}$Be} + \text{$^{1}_{1}$p}$) and is possible if the resulting nuclei have a total mass smaller than the mother nucleus one.

In the presence of LIV, however, spontaneous decay may happen even for nuclei which cannot spontaneously decay in the LI framework. The main difference with respect to the LI case is that now there is a certain energy threshold for the decay to occur. To analyze the kinematics of the process we use eq.~(\ref{ThresholdEquation}) by setting $\epsilon=0$, (i.e.~there is no low energy photon to interact with) since with this substitution the energy-momentum conservation eq.~(\ref{EnergyCons1}) still holds. The result is
\begin{equation} \label{pdecsol}
p^{n+2} = M^{n}_{Pl} \frac{\frac{m^{2}_{A',Z'}}{1-y} + \frac{m^{2}_{B,W}}{y} - m^{2}_{A,Z}}{\eta^{(n)}_{A,Z} - \eta^{(n)}_{A',Z'} (1-y)^{n+1} - \eta^{(n)}_{B,W} y^{n+1}}\,.
\end{equation}
This equation has either none or exactly one real and positive solution for $p$ depending on $y$. In the former case there is no spontaneous decay, similar to the LI situation, while in the latter case the (lower) spontaneous decay threshold $p^{dec}_{thr}$ is obtained by minimizing the result with respect to $y$.

Since the nuclei considered become unstable under such conditions, this kind of decay can be assumed to take place on a very short time scale.\footnote{In the absence of a proper theory of nuclei disruption it is impossible to reliably estimate this timescale. However, once above threshold this decay process is similar to the standard spontaneous decay, whose general timescales range from fractions of seconds to days or years. Moreover, the decay phase space grows with $p^{4}$ if $n=2$, hence the lifetime is expected to shorten with increasing energy.} This would mean that such particles cannot be observed on Earth. Analyzing the highest energy events under these aspects would therefore give information about the values of the LIV parameters.  

\subsection{Vacuum Cherenkov emission}

The Vacuum Cherenkov effect (VC), i.e.~the emission of a photon by a superluminal charged particle in vacuum (in the case here $^{A}_{Z}N \rightarrow \, ^{A}_{Z}N +\, \gamma$) is, just like the spontaneous decay described above, forbidden for the Lorentz invariant case due to energy-momentum conservation. In the LIV framework, however, it becomes possible and must be taken into account \cite{Jacobson:2002hd,Maccione:2009ju}.

Since the reaction time scale for VC is usually very small \cite{Jacobson2006150}, as low as $1\,\ns$ in the case of $10\,\TeV$ electrons, the mean free path for the reaction is negligibly small compared to intergalactic propagation. Therefore, if a nucleus is observed with a certain energy $\bar{E}$, the VC energy threshold must be larger than $\bar{E}$. Moreover, assuming that LIV is much smaller for photons than for nuclei \cite{PhysRevLett.100.021102,Maccione:2008iw,Galaverni:2008yj}, we consider only emission of soft photons. This is a valid assumption, as the phase space of the reaction just above threshold is large enough to warrant the short mean free path necessary to set constraints \cite{Maccione:2009ju}.

We compute again the threshold momentum by exploiting energy-momentum conservation
\begin{equation}
m_{A,Z}^{2} = \left( E_{A,Z}' + \epsilon_{VC} \right)^{2} - \left( \vec{p}_{A,Z}' + \vec{p}_{\gamma,VC} \right)^2\,.
\end{equation}
After performing a calculation similar to the one done for eq.~(\ref{EnergyCons1}), the momentum threshold is \cite{Jacobson:2002hd}
\begin{equation}
p_{thr,VC}^{n+2} = \frac{m_{A,Z}^{2} M_{Pl}^{n}}{(n+1)\eta_{A,Z}^{(n)}}
\end{equation}
which leads to the constraint
\begin{equation}
\eta_{A,Z}^{(n)} < \frac{m_{A,Z}^{2} M_{Pl}^{n}}{(n+1) \bar{E}^{n+2}}
\end{equation}
if a nucleus with energy $E \simeq \bar{E}$ is observed. 

\section{Results} \label{sec:Results}

We show here results obtained in the simplified case described in sec.~\ref{sec:DispRel}, considering $n=2$ with $\eta^{(2)}_{A,Z}=\frac{\eta}{A^2}$, $\eta$ being the LIV coefficient of a nucleon, and focusing on single nucleon emission. 

\begin{figure}[tbp]
  \centering
    \fbox{\includegraphics[scale=0.2572]{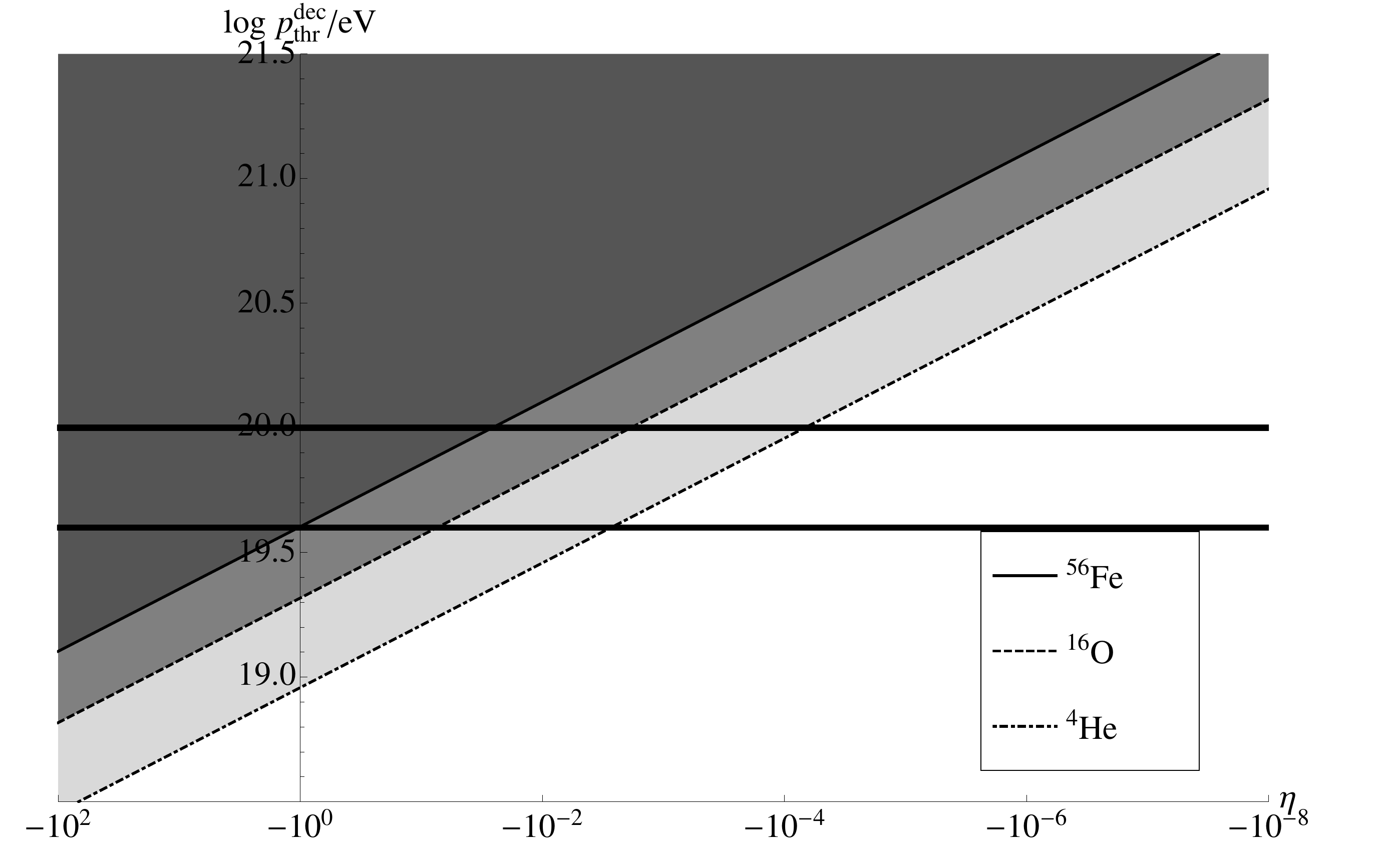}}
    \fbox{\includegraphics[scale=0.2572]{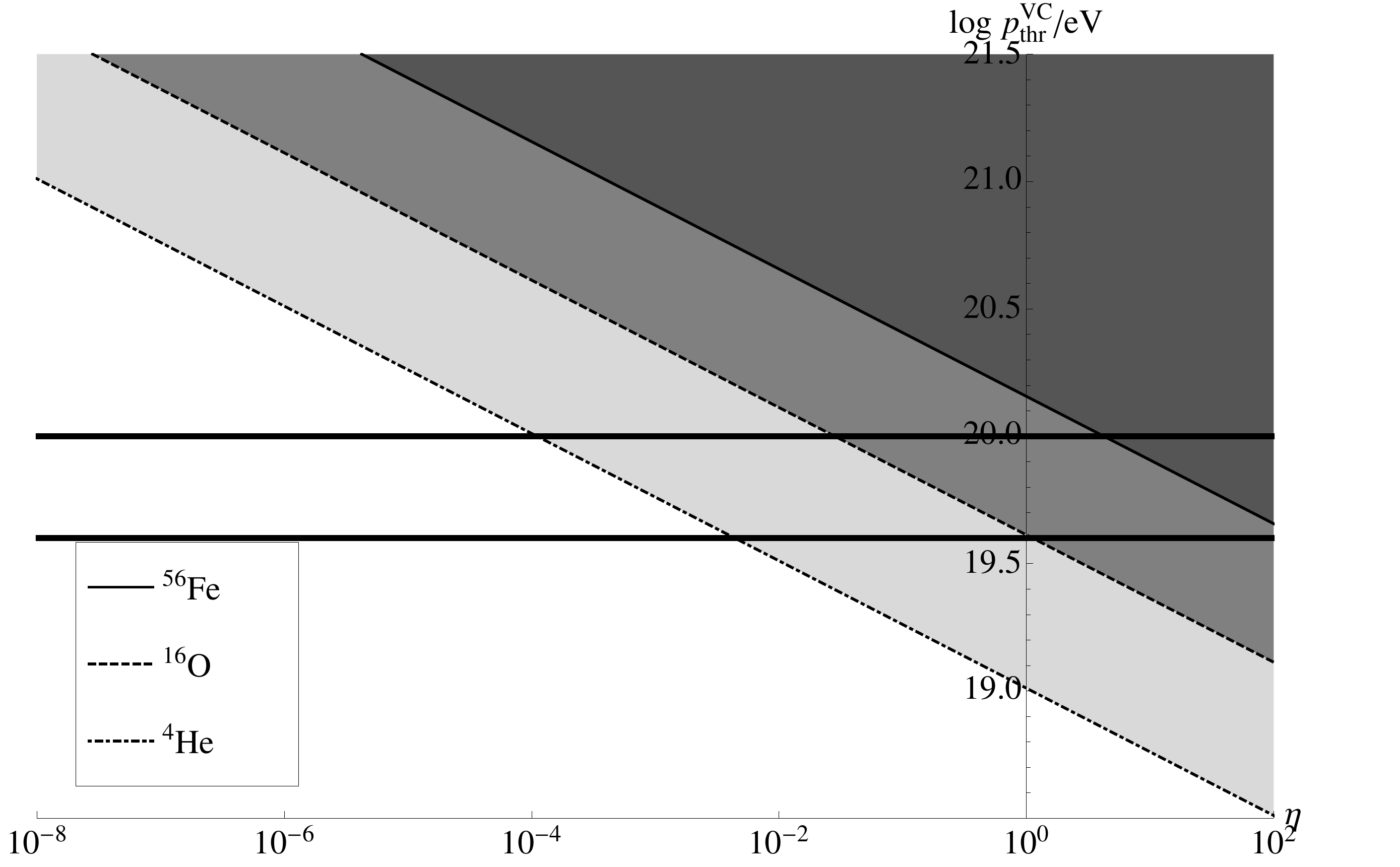}}
    \caption{Threshold momenta of spontaneous decay (left) and VC emission (right) for different elements. The respective shaded regions are excluded when the corresponding nucleus is observed at a given momentum value. Two of such observations are marked by the thick horizontal lines, taken at approx.~$10^{19.6}~\eV$ from \cite{AstropartPhys.33.151} and at $10^{20}~\eV$ from \cite{Abraham2010239}.} 
    \label{fig:DecVCthr}
\end{figure}

Assuming that current hints for a heavy composition at energies $E \sim 10^{19.6}~\eV$ \cite{Abraham:2010yv} may be confirmed in the future, and that some UHECR is observed up to $E \sim 10^{20}~\eV$ \cite{Abraham2010239}, one could place a first constraint on the absence of spontaneous decay for nuclei which cannot spontaneously decay without LIV as well. It will place a limit on $\eta<0$, because in this case the energy of the emitted nucleon is lowered with respect to the LI case until it ``compensates'' the binding energy of the nucleons in the initial nucleus in the energy-momentum conservation. We show in fig.~\ref{fig:DecVCthr} (left panel) the behavior of the threshold momentum with respect to $\eta<0$ for three representative nuclear species: $^{56}$Fe, $^{16}$O and $^{4}$He. If the UHECRs at the highest energies were Fe, then the limit is $\eta \gtrsim -1$ for $10^{19.6}~\eV$ (and $-3\times10^{-2}$ for $10^{20}~\eV$), while if they were lighter species, like He, then a remarkable constraint $\eta \gtrsim -3\times10^{-3}$ ($-7\times10^{-5}$) could be placed.
An upper limit for $\eta>0$ can instead be obtained from VC. This is illustrated in fig.~\ref{fig:DecVCthr} (right panel). Assuming UHECR to be mainly iron at the highest energies the constraint is given by $\eta \lesssim 2\times10^{2}$ for nuclei observed at $10^{19.6}~\eV$ (and $\eta \lesssim 4$ for $10^{20}~\eV$), while for helium it is $\eta \lesssim 4\times10^{-3}$ ($10^{-4}$). 

A distinct upper constraint could in principle be obtained by assuming that photodisintegration is at work in UHECRs and then computing the largest value of $\eta$ for which the reaction is kinematically allowed in the required energy range. The change of the threshold momentum due to LIV is, however, very small as it can be understood with the following reasoning: The threshold equation (\ref{ThresholdEquation}) gives a situation similar to the LI case for small $\eta$, i.e.~the threshold inelasticity is approximately $\displaystyle y_{thr} \approx y_{thr,LI} = \frac{m_{B,W}}{m_{A',Z'} + m_{B,W}}$. For large positive $\eta$ the first term in eq.~(\ref{ThresholdEquation}) becomes the dominating one, so the equation reduces to
\begin{equation} \label{ThresholdEquation_largeEta}
\eta \left( \frac{1}{A^{n}} - y^{n+1} - \frac{1}{(A')^{n}} (1-y)^{n+1} \right) \frac{p^{n+2}}{M^{n}_{Pl}} \approx 0\,.
\end{equation}
Solving this gives $y_{thr}$. For single nucleon emission it is $y_{thr} \approx \frac{1}{A}$ (which can be confirmed by plugging in), while for larger $B$ at $y_{thr} \approx \frac{B}{A}$ the left hand expression of eq.~(\ref{ThresholdEquation_largeEta}) is at least as close to zero as possible. Substituting this back into eq.~(\ref{ThresholdEquation}) and solving it for $p$ gives the lower threshold for large $\eta$. For $B=1$,
\begin{equation} \label{LowerThr}
\underset{\eta \rightarrow \infty}{\lim} \underline{p}_{thr} \approx \frac{A}{4 \epsilon}\left(\frac{m^{2}_{A-1,Z'}}{A-1}+m^{2}_{1,W}-\frac{m^{2}_{A,Z}}{A}\right).
\end{equation}
We checked {\it a posteriori}  that the approximation, eq.~(\ref{ThresholdEquation_largeEta}), is valid. Compared to the LI case $\underline{p}_{thr,LI} = \frac{1}{4\epsilon} \left( \left( m_{A'} + m_{1} \right)^{2} - m_{A}^{2} \right)$, the relative difference is
\begin{equation}
\frac{\Delta \underline{p}_{thr}}{\underline{p}_{thr,LI}} = \frac{\underline{p}_{thr} - \underline{p}_{thr,LI}}{\underline{p}_{thr,LI}} \approx (A-1) \frac{\left(\frac{m_{A'}}{A-1} - m_{1} \right)^{2}}{\left(m_{A-1}+m_{1}\right)^{2}-m_{A}^{2}}
\end{equation}
which is independent of $\eta$ in the limit of large $\eta$. For different elements we obtain $\frac{\Delta \underline{p}_{thr}}{\underline{p}_{thr,LI}} = 7\times10^{-5}$ for $\text{$^{4}$He}$, $2\times10^{-3}$ for $^{16}$O and $3\times10^{-3}$ for $\text{$^{56}$Fe}$. Hence an upper constraint cannot be set just by considering the threshold shift because the relative change does not exceed $3\times10^{-3}$ for elements up to iron and is therefore too small compared to the accuracy required to match the data.

On the other hand, statistical quantities related to the probability of the reaction to happen, i.e.~the different propagation lengths, can lead to interesting results. Now the question might arise how it is possible that these quantities are significantly affected by LIV while the thresholds are not. Indeed, the photon energy shift in the CR rest frame is linear with respect to $\eta$, thus making the phase space and with it the cross section at a given photon energy $\epsilon$ be dramatically affected by LIV.
\begin{figure}[tbp]
  \centering
    \fbox{\includegraphics[scale=0.35]{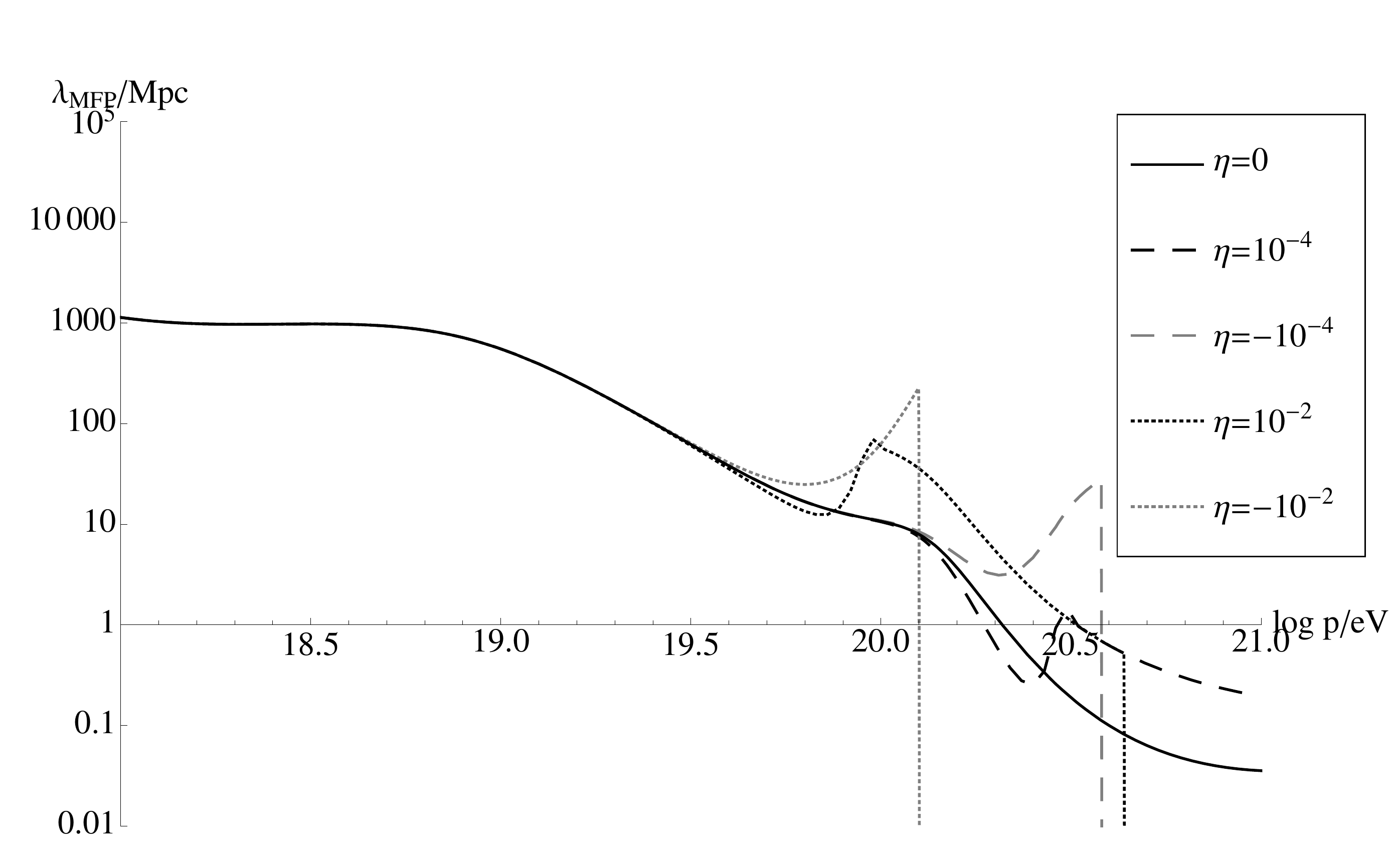}}
    \fbox{\includegraphics[scale=0.35]{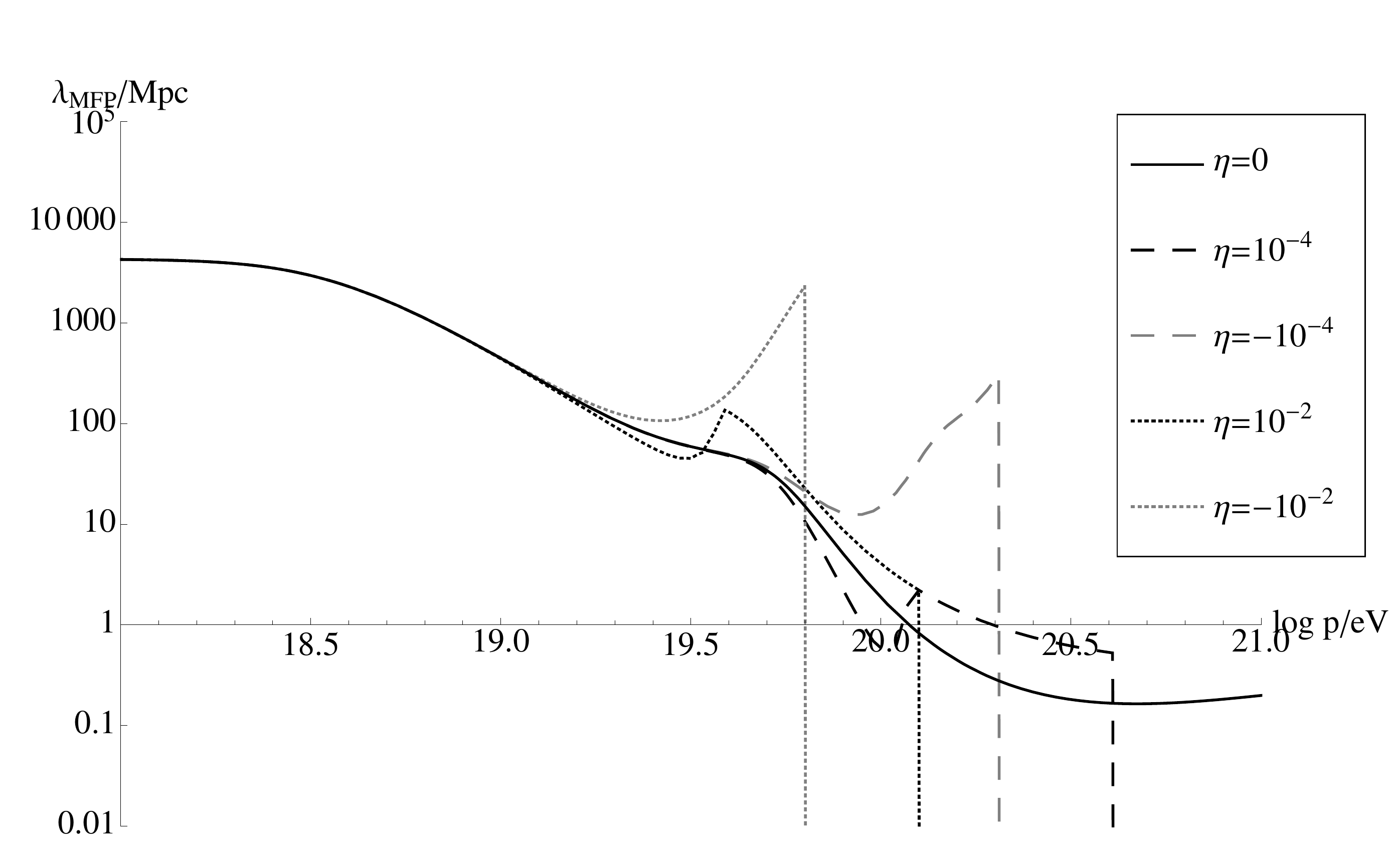}}
    \fbox{\includegraphics[scale=0.35]{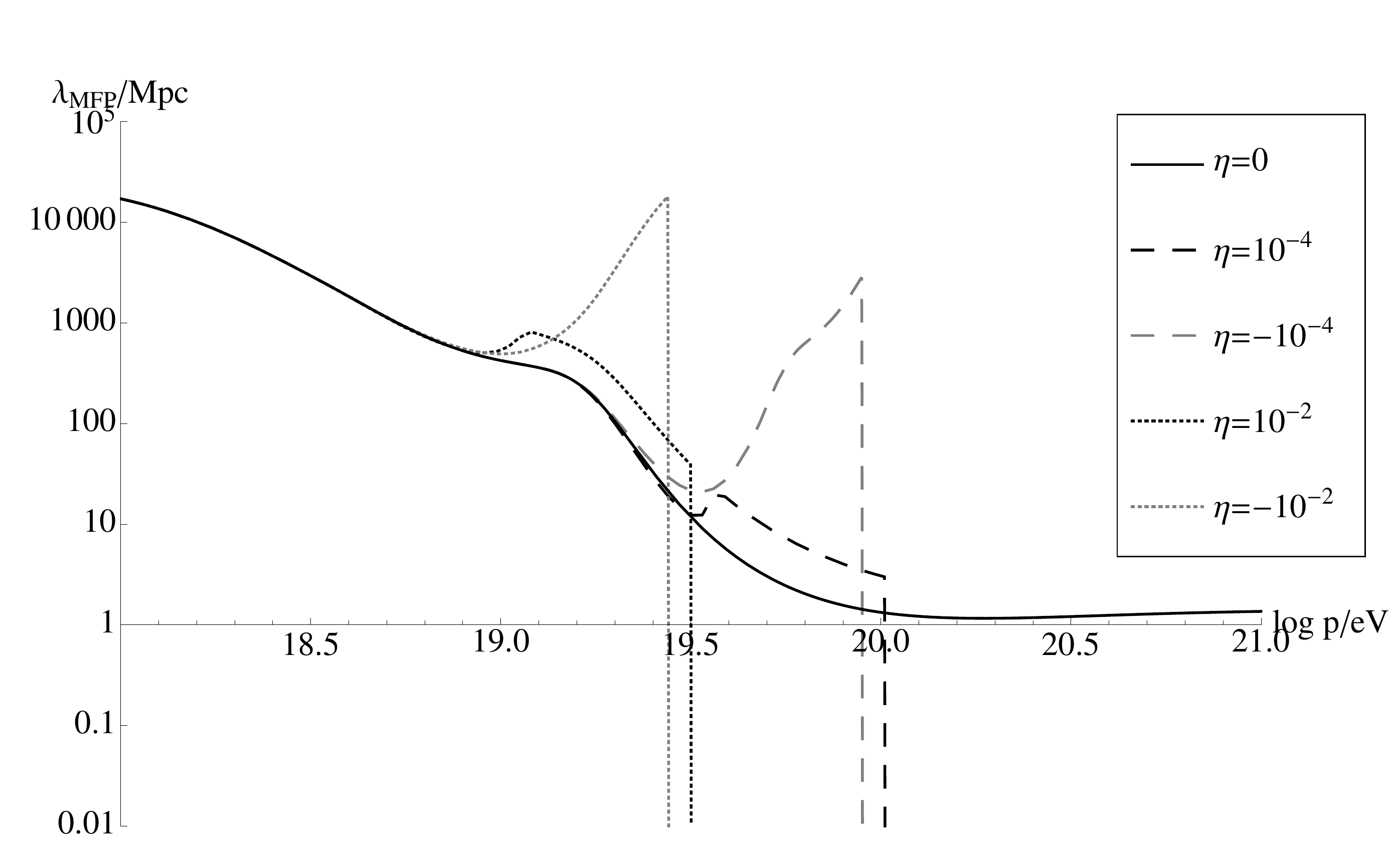}}
    \caption{The mean free path of photodisintegration in the single parameter model for $\text{$^{56}$Fe}$ (top), $\text{$^{16}$O}$ (middle) and $\text{$^{4}$He}$ (bottom).} 
    \label{fig:MFP}
\end{figure}

The mean free path for different elements and for different values of $\eta$ is shown in fig.~\ref{fig:MFP}. LIV introduces quite interesting deviations from the LI case. The mean free path differs by at least one order of magnitude at the highest energies. The deviation goes however in different directions: For positive LIV parameters it decreases with growing momentum down to distances below $1~\Mpc$ and then has a cutoff due to VC. On the other hand, for negative values it first increases (because of the reduced phase space), and then ends with a cutoff at the spontaneous decay threshold, i.e.~a very fast decrease of $\lambda_{MFP}$ down to cosmologically tiny values. 

This produces interesting consequences on the UHECR spectrum. First of all, assuming that the closest UHECR source lies at least at a few Mpc distance from the Earth, the UHECR spectrum should display a sharp cutoff of the heavy nuclei component corresponding to energies for which $\lambda_{MFP} \lesssim 1~\Mpc$. On the other hand, if UHECR nuclei are found at such large energies, with such a small MFP, then a UHECR source should be present at a distance closer than 1 Mpc from the Earth. This would produce a strong anisotropy of the flux towards the source direction at $E \gtrsim 10^{19.5}~\eV$. As it is clear from fig.~\ref{fig:MFP}, this reasoning has the potential to set a constraint $|\eta|< 10^{-2}$ even for $^{56}$Fe, once better data will be available. Given that the experimental data at such high energies still suffer from statistical and systematic uncertainties, we leave for future work the quantitative determination of the LIV UHECR spectrum and anisotropy, for which a full simulation of the propagation of heavy nuclei in the IGM, following the complete photodisintegration pattern, is needed.

\section{Conclusions} \label{sec:Conclusions}

Motivated by recent experimental results hinting at a significant presence of heavy nuclei in UHECRs, we analyzed the effects of a possible Lorentz invariance violation on the propagation of ultra high energy cosmic ray nuclei in the intergalactic medium. Relevant effects were found in the process of photodisintegration onto intergalactic medium background photon fields. Moreover, we studied also Lorentz invariance violation effects on the nuclear spontaneous decay, and on possible emission of Cherenkov radiation by superluminal charges in vacuum. While we did the computation of the main effects for a general modified dispersion relation, we obtained results only in the case of $O(E^{2}/\Mpl^{2})$ modified dispersion relations, assuming also that there is only one Lorentz invariance violation parameter controlling Lorentz invariance violation for all nuclear species, $\eta$.

A lower limit on $\eta$ can be set by requiring that nuclei, which are stable in the Lorentz invariant case, should not undergo spontaneous decay emitting single nucleons below energies of the order of $10^{19.5}$ to $10^{20}$~eV, at which a substantial fraction of heavy ultra high energy cosmic ray nuclei has been observed while an upper limit is given by an analogous argument for vacuum Cherenkov emission. 

Constraints obtained with these two techniques are summarized in Tab.~\ref{tab:SDVCConstraints}.
\begin{table}[tbp]
\centering
\begin{tabular}{|c||c|c|c|}
\hline
                   & $E_{max} = 10^{19.6}\,\eV$                      & $E_{max} = 10^{20}\,\eV$ \\
\hline
\hline
$\text{$^{4}$He}$  & $-3\times10^{-3} \lesssim \eta \lesssim 4\times10^{-3}$ & $-7\times10^{-5} \lesssim \eta \lesssim 1\times10^{-4}$ \\
\hline
$\text{$^{16}$O}$  & $-7\times10^{-2} \lesssim \eta \lesssim 1$   & $-2\times10^{-3} \lesssim \eta \lesssim 3\times10^{-2}$             \\
\hline
$\text{$^{56}$Fe}$ & $-1 \lesssim \eta \lesssim 200$                         & $-3\times10^{-2} \lesssim \eta \lesssim 4$              \\
\hline
\end{tabular}
\caption{Constraints given by spontaneous decay and vacuum Cherenkov emission if a nucleus of the species given in the left column with energy $E_{max}$ is observed.}
\label{tab:SDVCConstraints}
\end{table}
These constraints are to be compared to the results obtained assuming ultra high energy cosmic rays to be only protons \cite{Maccione:2009ju}. In \cite{Maccione:2009ju} it was found, for the case of pure proton composition, $-10^{-3} \lesssim \eta \lesssim 10^{-6}$. As it can be seen, our constraints improve when lighter nuclei are considered, and agree well with the ones presented in \cite{Maccione:2009ju} when extrapolated to pure proton composition. We remark however that the recent findings of the presence of heavy nuclei in the ultra high energy cosmic ray spectrum might in fact invalidate the hypothesis of pure proton composition adopted in \cite{Maccione:2009ju}.

It is notable that we did not derive bounds coming directly from Lorentz invariance violation effects in photodisintegration. On the one hand, present experimental data on the chemical composition are still affected by systematics. On the other hand, to actually compute the theoretically expected spectrum and chemical composition in the presence of violations of Lorentz invariance requires strong computational efforts. Therefore, at present it is not possible to derive robust limits from the spectral information. Nevertheless, we derived expected order-of-magnitude limits assuming that photodisintegration be at work at energies of order $10^{19.6}~\eV$ for all nuclear species, and we leave a more detailed evaluation of the constraints for future work.

\section*{Acknowledgments}
We gratefully thank A.~Grillo and S.~Liberati for reading the draft and providing useful comments. This work was supported by the Deutsche Forschungsgemeinschaft through the collaborative research centre SFB 676. LM and GS acknowledge support from the State of Hamburg, through the Collaborative Research program ``Connecting Particles with the Cosmos'' within the framework of the LandesExzellenzInitiative (LEXI).

\bibliography{bib_updated}

\end{document}